\documentclass[letter, 11pt]{revtex4-1}
\usepackage{epsfig,epstopdf,amsmath,amssymb,graphicx,subfigure,verbatim,natbib}

\usepackage{calligra}
\DeclareMathAlphabet{\mathcalligra}{T1}{calligra}{m}{n}
\DeclareFontShape{T1}{calligra}{m}{n}{<->s*[2.2]callig15}{}

\begin{document}

\author{John A. Brehm}
\email{brehmj@sas.upenn.edu}
\affiliation{The Makineni Theoretical Laboratories, Department of Chemistry, University of Pennsylvania, 231 S. 34th Street Philadelphia, Pennsylvania  19104-6323}

\author{Joseph W. Bennett}
\affiliation{The Makineni Theoretical Laboratories, Department of Chemistry, University of Pennsylvania, 231 S. 34th Street Philadelphia, Pennsylvania 19104-6323}  

\author{Michael Rutenberg Schoenberg}
\affiliation{The Makineni Theoretical Laboratories, Department of Chemistry, University of Pennsylvania, 231 S. 34th Street Philadelphia, Pennsylvania 19104-6323}

\author{Ilya Grinberg}
\affiliation{The Makineni Theoretical Laboratories, Department of Chemistry, University of Pennsylvania, 231 S. 34th Street Philadelphia, Pennsylvania 19104-6323}

\author{Andrew M. Rappe}
\email{rappe@sas.upenn.edu}
\affiliation{The Makineni Theoretical Laboratories, Department of Chemistry, University of Pennsylvania, 231 S. 34th Street Philadelphia, Pennsylvania 19104-6323}

\date{\today}

\begin{abstract}

We use first-principles density functional theory within the local density approximation  to ascertain the ground state structure of real and theoretical compounds with the formula $AB$S$_\text{3}$  ($A$ = K, Rb, Cs, Ca, Sr, Ba, Tl, Sn, Pb, and Bi; and $B$ = Sc, Y, Ti, Zr, V, and Nb) under the constraint that $B$ must have a $d$$^\text{$0$}$ electronic configuration. Our findings indicate that none of these $AB$ combinations prefer a perovskite ground state with corner-sharing $B$S$_\text{6}$ octahedra, but that they prefer phases with either edge- or face-sharing motifs.   Further, a simple two-dimensional structure field map created from $A$ and $B$ ionic radii provides a neat demarcation between combinations preferring face-sharing versus edge-sharing phases for most of these combinations. We then show that by modifying the common Goldschmidt tolerance factor with a multiplicative term based on the electronegativity difference between $A$ and S, the demarcation between predicted edge-sharing and face-sharing ground state phases is enhanced. We also demonstrate that, by calculating the free energy contribution of phonons,  some of these compounds may assume multiple phases as synthesis temperatures are altered, or as ambient temperatures rise or fall.    

\end{abstract}

\title{The Structural Diversity of $AB$S$_\text{3}$ Compounds with $d$$^\text{$0$}$ Electronic Configuration for the $B$-cation}
\maketitle

\section{Introduction}

In a key work, Muller and Roy used the crystal chemistry method of cation-anion coordination to categorize many of the compounds found experimentally in the major ternary structural families $A$$_\text{2}$$B$$X$$_\text{4}$, $AB$$_\text{2}$$X$$_\text{4}$, and $ABX$$_\text{3}$ known at the time of its publication in 1974.\cite{Muller74p1}  In their analysis of $ABX$$_\text{3}$ compounds, they constructed structure field maps for those compounds with anions $X$ = O, F, and Cl.  These maps plot structure as a function of $A$ and $B$ ionic radii and often lead to regions on the diagrams where only certain phases have been realized experimentally.  From these maps, the structures for other $A$ and $B$ pairs can be predicted.  Absent from their analysis is any structure field map of $AB$S$_\text{3}$ compounds.  Indeed, very few of the compounds listed in their $ABX$$_\text{3}$ section have $X$ = S, and, for those that do, some of these have phases that were reported as not known with certainty ($e.$ $g.$ the compounds CaZrS$_\text{3}$ and SrZrS$_\text{3}$ synthesized by Clearfield\cite{Clearfield63p135}).  

Since Muller and Roy's work, the number of synthesized $AB$S$_\text{3}$ compounds has increased substantially; these
show a distribution of structural motifs that is in stark contrast to their  $AB$O$_\text{3}$  analogs.  Most $AB$O$_\text{3}$ compounds show networks of corner-sharing  $B$O$_\text{6}$ octahedra and are commonly called perovskites.  Several form in the ilmenite phase, in which layers of edge-connected $A$O$_6$ octahedra are connected by faces and corners to layers of edge-connected $B$O$_6$ octahedra.  As well several other $AB$O$_\text{3}$ do not have any corner-, edge-, or face- sharing designation, but are instead distinguished  by $B$O$_3$  $B$ = B, C, N, S, Cl, Br,  and I anionic complexes.   Pyroxenes are also a less common, but still noteworthy, subclass of $AB$O$_\text{3}$ types, in which $B$O$_4$ tetrahedra are corner-connected. Only a couple of $AB$O$_\text{3}$  have been found in phases with solely face-sharing or edge-sharing $B$O$_\text{6}$ octahedral motifs.  In contrast,  $AB$S$_\text{3}$ compounds are observed with networks of either solely corner-, edge-, or face-sharing $B$S$_6$ octahedral motifs.  For example, BaZrS$_\text{3}$ and CaZrS$_\text{3}$ form as corner-sharing perovskites; PbZrS$_\text{3}$ and TlTaS$_\text{3}$ form the edge-sharing NH$_\text{4}$CdCl$_\text{3}$ phase; and  BaTiS$_\text{3}$, BaVS$_\text{3}$, and BaNbS$_\text{3}$ form face-sharing structures.  Further, they are not known to form pyroxenes and there are only two instances listed in FIZ Karlsruhe ICSD database in which $AB$S$_3$ have anionic complexes with the $B$ mentioned above in $AB$O$_3$:  RbBS$_3$ and TlBS$_3$.\cite{Belsky02p364,FIZ13p1}  They do not form the layered ilmenite phase either, but there are near stoichiometric compositions of $AB$S$_\text{3}$ (denoted as misfit sulfides) with sheets of edge-sharing $B$S$_\text{6}$ octahedra sandwiching incommensurate rock salt-like $A$S layers.   Two examples of misfit layered compounds are (SnS)$_\text{1.12}$TiS$_\text{2}$ and (PbS)$_\text{1.18}$TiS$_\text{2}$.

Unlike their  $AB$O$_\text{3}$ analogs, $AB$S$_\text{3}$ compounds are not neatly classified by the Goldschmidt tolerance factor,\cite{Goldschmidt26p477}

\begin{eqnarray}
t = \frac{r_A + r_X}{\sqrt{2}(r_B+r_X)}   
\end{eqnarray}

\noindent where the various $r$ represent the ionic radii of the constituent species.  A $t$ = 1 indicates ideal packing in the cubic perovskite structure.  As shown by Woodward for $AB$O$_\text{3}$, the corner-sharing perovskite phase is stable for \text{$\approx$}0.95  {\textless } $t$  {\textless } 1.05,  with most octahedral tilts being observed for $t$  {\textless } 1, and most untilted structures being realized for $t$  {\textgreater } 1.\cite{Woodward97p44}  Coupled corner- and face-sharing phases ({\em e. g.} SrMnO$_\text{3}$ and BaRuO$_\text{3}$) begin to form when $t$ {\textgreater } 1.04, and completely face-sharing phases with no corner-sharing character form when $t$ {\textgreater } 1.10 ($e.$ $g.$ BaMnO$_\text{3}$).  The solely edge-sharing phase is rare in $AB$O$_\text{3}$  according to Goodenough, and he lists just a single case in his extensive review of $AB$O$_\text{3}$ compounds:  RbNbO$_\text{3}$ with $t$ = 1.085.\cite{Goodenough04p1915}  The ilmenites form with $t$ {\textless } $\approx$ 0.8. Except for one or two cases, the pyroxenes and those $AB$O$_3$ compounds with $B$O$_3$ anionic complexes form with $t$ greater than those of the corner-sharing perovskites.  Furthermore, except for a few compounds with $B$ = S or Cl, those compounds with anionic complexes having $B$ = B, C, N, S, or Cl, have $t$ factors strictly greater than those of solely face-sharing structures. However, in the case of $AB$S$_\text{3}$, just for the compounds listed above, overlapping ranges are obtained:  0.88  {\textless } $t$  {\textless } 0.95 for corner-sharing structures, 0.92  {\textless } $t$  {\textless } 1.01 for edge-sharing structures, and 0.98  {\textless } $t$  {\textless } 1.03 for face-sharing structures.   

In the early 1980s, Pettifor developed structure field maps in a different way from Muller and Roy.  Instead of using the ionic radii for the abscissa and the ordinate, he defined a chemical scale based on the results of phase groupings of 574 binary compounds.\cite{Pettifor84p31}  The elements, from hydrogen through the actinides, were scaled in such a manner that the resulting list also mirrored, to a large extent, an ordering of the elements by electronegativity.\cite{Ferro96p205}   In 1988, he applied his mapping method to various ternary formula families including $AB$S$_\text{3}$ compounds.\cite{Pettifor88p675}  However, unlike the Muller and Roy maps of $ABX$$_\text{3}$ $X$=O, F, and Cl, Pettifor's map did not lead to a good demarcation between edge-sharing compounds and corner-sharing ones.   
Furthermore, if edge-sharing compounds not included in his figure (such as PbSnS$_\text{3}$, BaSnS$_\text{3}$, PbZrS$_\text{3}$, and SnZrS$_\text{3}$) are also considered, demarcations between phases of different motifs becomes even more blurred.   Finally, the discovery of the stable edge-sharing phase of SrZrS$_\text{3}$ by Lee {\em et al.}\cite{Lee05p1049} in 2005 also diminishes the distinction between edge- and corner-sharing regions of his map.

In the current paper, we investigate the disagreement between $t$ factor expectations and experimental phase results in $AB$S$_\text{3}$ and develop a methodology for predicting the ground state structures of $AB$S$_\text{3}$ compounds and energetically competitive crystal structures that could be reasonably stabilized.   We also calculate the local density approximation (LDA) band gap for the ground  state phase and these alternate phases to highlight the structure-property differences.

\section{Methodology}

In order to determine the ground state structural tendencies of $AB$S$_\text{3}$ compounds, we first construct a sample subset of 20 compounds.  The $A$-sites considered are the Group 1 elements K, Rb, and Cs, the Group 2 elements Ca, Sr, and Ba, the Group 13 element Tl, the Group 14 elements Sn and Pb, and the  Group 15 element Bi.  To focus the study, Period 4 and 5 $B$-site cations are chosen such that the electron configuration is $d$$^\text{$0$}$: Sc, Y, Ti, Zr, V, and Nb.
While no combinations of Group 1 or Tl $A$-sites for $AB$S$_\text{3}$  are known to exist,  (except for TlTaS$_\text{3}$ with Ta outside the scope of this study), they are considered as interesting extensions to various Ba$B$S$_\text{3}$ that do exist:  all  are as large or larger than Ba$^\text{2+}$ in 12-fold coordination.   Further, $ABX$$_\text{3}$ oxides and halides with $A$ = K, Rb, and Cs do exist.  
For each of these compounds, we then arrange the atoms into 22 phases that $ABX$$_\text{3}$ compounds are known to assume.  Then, using density functional theory (DFT) within the LDA approximation, we calculate the relative energy of each phase with the ABINIT computing package.\cite{Gonze02p478} 

The set of 22 phases chosen includes the most common experimentally found corner-, edge-, and face-sharing $BX$$_\text{6}$ octahedral structures. The corner-sharing arrangements chosen are the cubic $Pm3m$, the tetragonal $P4mm$, the low temperature $R3mR$ BaTiO$_\text{3}$ phase and two tilt systems denoted by Glazer's naming scheme:\cite{Glazer72p3384} the common $a$$^\text{$+$}$$b$$^\text{$-$}$$b$$^\text{$-$}$ $Pnma$  and the low temperature $a$$^\text{$0$}$$a$$^\text{$0$}$$c$$^\text{$-$}$  $I4/mcm$  of SrTiO$_\text{3}$.  

For edge-sharing systems, we consider four phases.  The first, the commonly found NH$_\text{4}$CdCl$_\text{3}$ $Pnma$ phase, has as its defining pattern double columns of edge-sharing $B$S$_\text{6}$ octahedra, with each octahedron sharing edges with four others. The second is the $Pna2_1$ phase, which differs from this first phase in that atoms are displaced from high symmetry positions preserving a screw axis symmetry along the direction of the columns.    The third edge-sharing phase is similar to the second, but displacements of atoms in the plane perpendicular to the screw axis are permitted.  This phase is well known for the family of YScS$_\text{3}$ $Pna2_1$ structures, a group of compounds with lanthanide element $A$ sites in which the edge-sharing occurs for $A$S$_\text{6}$ prisms and the $B$S$_\text{6}$ are corner-connected.  In order to distinguish between these two phases, we term them  $E\_Pna2_1$ and  $C\_Pna2_1$ respectively, with the $E$ (edge) and $C$ (corner) indicating the 
connectivity of the $B$S$_\text{6}$ octahedra.  The remaining edge-sharing phase is a very low symmetry $P1$ phase found for RbNbO$_\text{3}$.

Four of the face-sharing phases we consider are based on the research of Fagot {\em et al.}\ and Ghedira {\em et al}.:\cite{Fagot05p718,Ghedira81p1491}  the $Cmcm$, the $C222_1$,  the $Cmc2_1$, and the $P6_3/mmc$.  Like all face-sharing phases, they have separated single columns of  face-sharing octahedra.  The first three have an orthorhombic lattice.  $Cmcm$ has  $B$ cations occupying high symmetry coordinates (0 and 0.5) in all three Cartesian directions leading to collinear $B$ cations in the columnar direction; in  $C222_1$, the $B$ cations are non-collinear and zigzag about one of the directions perpendicular to the columns; and in $Cmc2_1$, the $B$ cations zigzag in both directions  perpendicular to the columns.  The $P6_3/mmc$ phase is similar to the  $Cmcm$ phase in that both  have two mirror planes and one glide plane.  They differ in that $Cmcm$ is an orthorhombic crystal system with  a base-centered Bravais lattice, while $P6_3/mmc$ is a hexagonal crystal system with a simple Bravais lattice.  $P6_3/mmc$ is classified as a minimal non-isomorphic subgroup of $Cmcm$.   Two other hexagonal face-sharing phases, the $P6_3cm$ and $P6_3mc$,  are also evaluated.  $P6_3mc$ differs from $P6_3/mmc$ in that the former allows $B$ cation shifts from high symmetry positions in the column direction. For the $P6_3cm$ phase, there are two distinct sets of  columns of face-sharing octahedra which  are offset by a 1/4 unit vector in the column direction.  Thus, in total, six face-sharing phases are evaluated.

The remaining seven phases considered are either one of three types of mixed motif phases, or a corner-sharing tetrahedral phase.
Three mixed motif face-sharing and corner-sharing phases are evaluated in this study and are labelled based on the fraction of face-sharing octahedra per unit cell as $2/3$, $1/2$, and $1/3$.  They are most easily visualized by considering their projections on the (110) plane:  the first consists of stacks of three face-sharing octahedra joined at a corner; the second consists of stacks of two face-sharing octahedra joined at a corner; and the third consists of alternating stacks of two face-sharing octahedra sharing a corner with a single octahedron.  Respectively, these phases are known by their structure type names as BaRuO$_\text{3}$, BaMnO$_\text{3}$, and BaFeO$_\text{2+$x$}$.  A second type of mixed motif phase, one with mixed edge- and corner-sharing connectivity, is also considered.  The $Cmcm$ phase of the compound CaIrO$_\text{3}$,  (which is proposed to exist under high pressures for MgSiO$_\text{3}$),\cite{Murakami04p855} and its subgroup $Cmc2_1$, (in which atoms are no longer confined to high symmetry positions along the $z$-axis),\cite{Martin07p1912} have this motif and are respectively designated $MM\_Cmcm$ and $MM\_Cmc2_1$, with $MM$ signifying ``mixed motif''. This phase is characterized by planes of $B$S$_\text{6}$ octahedra, in which the octahedra are connected by edges in one direction in the plane, and by corners in the other planar direction. The last mixed motif phase considered is the ilmenite.  This phase, most often characterized by small $A$ and $B$ $d$-metal elements in $AB$O$_3$, has alternating layers of edge-sharing $A$O$_6$ octahedra and edge-sharing $B$O$_6$ octahedra.  The layers are connected by both face- and corner- sharing octahedra.  This phase is designated $MM\_Ilmen$.  Finally, a pyroxene $Pbcm$ phase consisting of single columns of zig-zag corner-sharing tetrahedra is included, as all compounds with the $A$VO$_\text{3}$ ($A$ = K, Rb, Cs, and Tl) chemical formula assume this structure.

The elements used in the study are represented in the DFT calculations by non-local,\cite{Ramer99p12471} norm-conserving optimized pseudopotentials\cite{Rappe90p1227} created with OPIUM.\cite{OpiumSourceforgenet11p1}   The plane wave cutoff energy used for both the pseudopotentials and the DFT calculations is 50 Ry.  We vary the Monkhorst Pack (MP) grid\cite{Monkhorst76p5188} depending on the size of the unit cell.  A k-point mesh of 16 is used along the reciprocal lattice directions for which the lattice parameter is  \text{$\approx$}5 \AA; 8 if it is \text{$\approx$}10 \AA, and 4 if it is  \text{$\approx$}20 \AA.  All hexagonal phases use grid shifts of 0$\times$0$\times$0.5; all others incorporate a shift of  0.5$\times$0.5$\times$0.5.  Where an LDA band gap calculation is required on a relaxed structure for a particular compound, an unshifted MP grid is used. We consider a structure to be relaxed when successive self-consistent iterations yield total energy differences  of  less than 10$^{\rm{-8}}$ Ha/unit cell and and atomic forces less than 10$^{\rm{-4}}$ Ha/Bohr.

For each of  the lowest energy phases of the 20 compounds, and for those phases nearest to them in terms of relative energy, we obtain the entropy contribution to the free energy and assess compound stability  by calculating the phonon normal mode frequencies at the $\Gamma$-point and then using  the equation:

\begin{eqnarray}
F_{\rm{vib, solid}} = \sum_{ s=1}^{3N}\left\{ \frac{\hbar\omega_s}{2} + k_BT{\rm{ln}}\Big[1 - {\rm{exp}}\Big(\frac{-\hbar\omega_s}{k_BT}\Big)\Big]\right\}
\end{eqnarray}
\noindent where $N$ represents the number of atoms in the system, $\omega$$_s$ represents a $\Gamma$-point normal mode frequency in the harmonic approximation, $k$$_B$ is the Boltzmann constant, and $T$ is temperature.

For the full set of 20 $A$-$B$ combinations, we develop two structure field maps  to elucidate sulfide structural preferences with respect to $A$ and $B$ cation sizes:  one following the method of Pettifor; the other the method of Muller and Roy.  In the case of the Muller and Roy type map, we combine the originally separated field maps for $A$$^\text{1+}$$B$$^\text{5+}$$X$$_\text{3}$, $A$$^\text{2+}$$B$$^\text{4+}$$X$$_\text{3}$, and $A$$^\text{3+}$$B$$^\text{3+}$$X$$_\text{3}$ into one plot for brevity.   For all ionic sizes, except  Sn$^\text{2+}$, we use the data found in Seshadri\cite{Seshadri11p1} and Shannon.\cite{Shannon76p751} We use the value of 1.4 $\text{\AA }$ for the ionic radius for Sn$^\text{2+}$ which was calculated  by Bennett {\em et al.}\cite{Bennett11p144112}   We use  12-fold coordination radii for $A$, and six-fold coordination radii for $B$ and S.  Using these radii, we then reassess the Goldschmidt $t$ factor in light of the preferred phases found.

\section{Results}

\begin{figure}[h]
\vspace{-20pt}
\includegraphics[scale = 0.5]{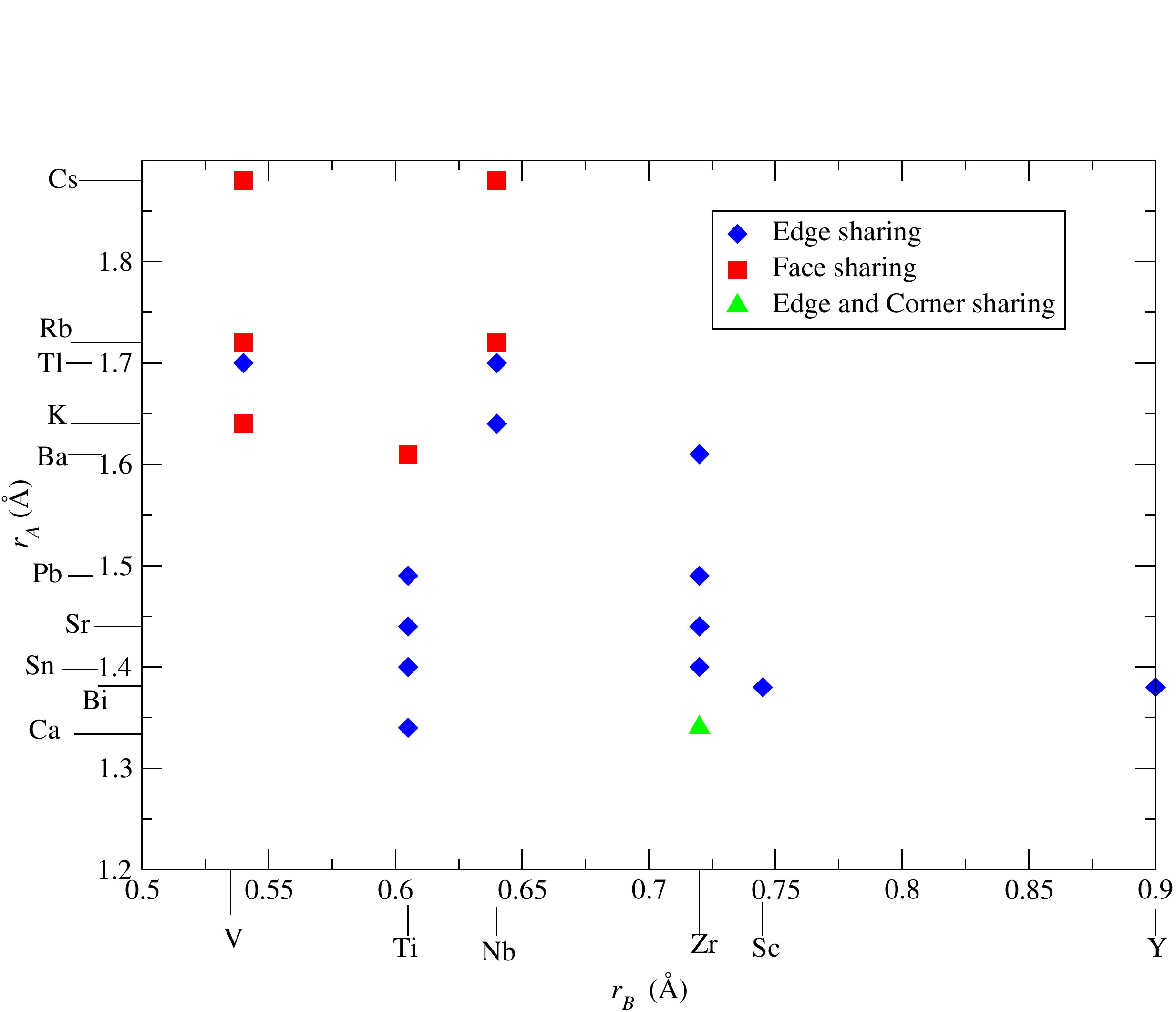}
\vspace{-10pt}
\caption{\label{structmap} Structure field map of ground state $AB$S$_\text{3}$ structures with various $B$S$_\text{6}$  octahedral motifs. All $r_A$  assume a coordination number of 12; all $r_B$  assume a six-fold coordination. }
\end{figure}

\begin{figure}[h]
\vspace{-10pt}
\includegraphics[scale = 0.5]{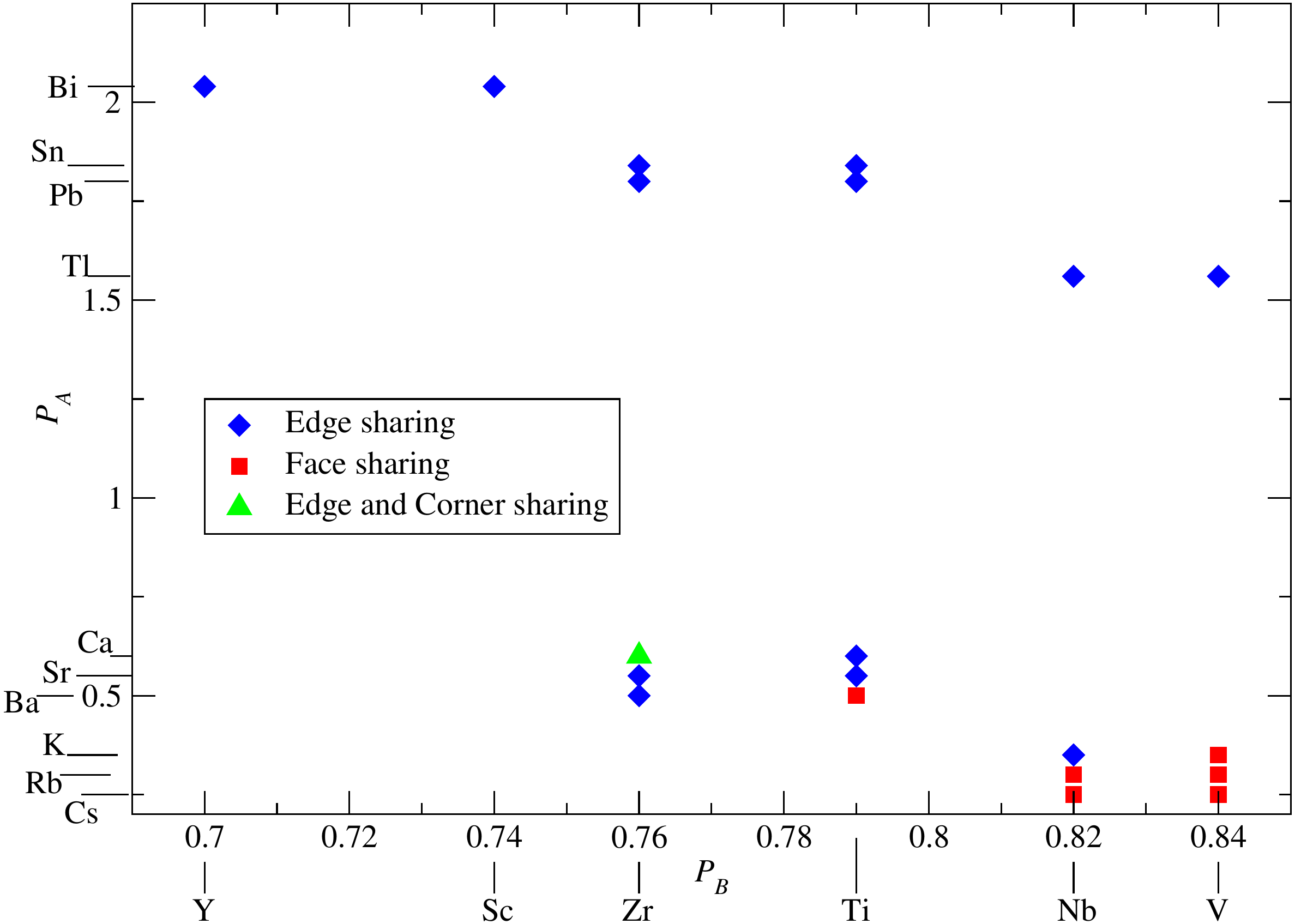}
\vspace{-10pt}
\caption{\label{structmap2} Pettifor chemical scale structure field map of ground state $AB$S$_\text{3}$ structures with various  $B$S$_\text{6}$  octahedral motifs.  $P$$_\text{A}$ and $P$$_\text{B}$ represent the values of the scale assigned to each element.  The value assigned to each element approximates its electronegativity.  Further details as to the construction of this scale can be found in \cite{Ferro96p205}.}
\end{figure}

The calculated ground state phases for the 20 $AB$S$_\text{3}$ compounds are plotted on the Muller and Roy ionic radii type structure field map in Figure \ref{structmap} and the Pettifor type structure field map in Figure \ref{structmap2}.   
With the exceptions of two  $A$ = Tl based compounds, the Muller and Roy type map demonstrates a well-defined demarcation  between the face-sharing and edge-sharing ground state structures.  As with the $AB$O$_\text{3}$ structure maps of Muller and Roy, the face-sharing ground state phase is found only in the regions of large $r_A$ and small $r_B$. Significantly, no pure corner-sharing perovskite is calculated to be the ground state phase.  Only the predicted ground state of the mixed motif corner- and edge-sharing $MM\_Cmc2_1$ phase for CaZrS$_\text{3}$ has any corner-sharing character. Even for this compound, the pure edge-sharing phase is slightly favored over the pure corner-sharing phase by 0.007 eV/20-atom unit cell.   The Pettifor map improves on the Muller and Roy type map in the sense that plotting the ground state structure in the Pettifor map leads to no overlap between the face- and edge-sharing ground state structures.   However, in the Pettifor map (Figure \ref{structmap2}) the mixed motif corner- and edge-sharing  CaZrS$_\text{3}$ falls within the edge-sharing region.

\begin{table*}
\caption{\label{rankme} Ranking of phases by $\Delta$$E$, the total energy per 20-atom cell for the ten $AB$S$_\text{3}$ $B$ = Ti and Zr compounds described in the text. In addition to the ground state energy phase, phases with different $B$S$_\text{6}$ motifs are also presented if they are within \text{$\approx$}1 eV per 20-atom cell of the ground state.  C = corner-sharing, E = edge-sharing, E/C = edge- and corner-sharing, and F = face-sharing. The number in the parentheses of the $\Delta$$E$ column is the difference in energy when the zero point of energy (ZPE) obtained from the phonon calculation is included. Structures  that were found to be unstable due to negative phonons at the $\Gamma$-point are indicated with NP. The $T_{ trans}$ column indicates the temperature at which the different structures have the same free energy relative to the ground state, and the system is predicted to undergo a phase transition. For some phases, there is no transition temperature, labelled NT.  LDA band gaps ($E_g$) are listed and labeled with an I/D = indirect/direct. Please refer to the Methodology Section for space group nomenclature.}
\scalebox{0.79}{\begin{ruledtabular}
\begin{tabular}{|c|cccc|ccccc|} 
$A$ & Motif &  $\Delta$$E$ (+ZPE) &  $T_{ trans}$ &$E_g$ & & Motif &  $\Delta$$E$ (+ZPE)  & $T_{ trans}$ & $E_g$  \\
 & (Sp. Grp.) &  (eV/unit cell) &  ( K) &  (eV) & & (Sp. Grp) &  (eV/unit cell)  & ( K) &  (eV)  \\
\hline
& \multicolumn{9}{|c|} {$A$$^\text{$2+$}$$B$$^\text{$4+$}$} \\ 
\hline
 &  \multicolumn{4}{c|} {Ti} &  & \multicolumn{4}{c|} {Zr} \\
\hline
Sn	&	E($ Pnma$)	&	0 (0)	&		&	0	& &	E ($Pnma$)	&	0 (0)	&		&	0.65 (D)  \\
	&	E/C ($MM\_Cmcm$)	&	0 (NP)	&	“------”	&	0	& &	E/C ($MM\_Cmc2_1$)	&	0.54 (0.49)	&	1075	&	0.55 (I)  \\
	&	C ($Pnma$)	&	0.87 (0.84)	&	NT	&	0	& &	C($Pnma$)	&	1.03 (0.97)	&	2475	&	0.24 (D)  \\
\hline																		
Pb	&	E ($Pnma$)			&	0 (0)	          &		&	0.16 (D)	& &	E ($Pnma$)	                     &	0 (0)		&			&	0.88 (D)  \\
	&	E/C ($MM\_Cmcm$)      & 0.26 (0.23)	&	1150	&	0	          & &	E/C ($MM\_Cmcm$)	&	0.167 (NP)	&	“------”	&	0.76 (D)   \\
	&	C($Pnma$)			&	0.56 (0.55)	&	NT	&	0	& &	C($Pnma$)			&	0.48 (0.44)	&	2180		&	0.53 (I)   \\
	&	F/E/C($MM\_Ilmen)$)			&	0.76 (0.79)	&	\textgreater4000	&	0.32 (I)	& &	F/E/C($MM\_Ilmen)$)				&	0.90 (0.89)	&	3750		&	0.65 (I)   \\
\hline																		
Ca	&	E ($Pna2_1$)	&	0 (0)	&		&	0	& &	E/C ($MM\_Cmc2_1$)	&	0 (0)	&		&	0.10 (D)  \\
	&	E/C ($MM\_Cmc2_1$)	&	0.11 (0.09)	&	305	&	0	& &	E ($Pnma$)	&	0.23 (0.24)	&	NT	&	0.27 (D)  \\
	&	C ($Pnma$)	&	0.16 (NP)	&	“-----”	&	0.14 (D)	& &	C ($Pnma$)	&	0.23 (0.25)	&	NT	&	0.96 (D)  \\
	&				&				&			&			& &F/E/C($MM\_Ilmen)$)		&	0.85 (0.89)	&	NT	&	1.67 (I)  \\
\hline	
Sr	&	E ($Pna2_1$)	&	0 (0)	&		&	0.15 (I)	& &	E ($Pnma$)	&	0	(0)	&		&	0.24 (D)  \\
	&	C ($Pna2_1$)	&	0.37 (0.38)	&	NT	&	0.54 (D)	& &	E/C ($MM\_Cmc2_1$)	&	0.16 (0.14)	&	1100	&	0.23 (D) \\
	&	E/C ($MM\_Cmc2_1$)	&	0.51 (0.50)	&	NT	&	0	& &	C ($Pnma$)	&	0.29 (0.27)	&	2050	&	0.96 (D) \\
\hline	 		 			 		 		 		 			 		 	
Ba	&	F ($C222_1$)	&	0 (0)	&		&	0	& &	E ($Pnma$)	&	0 (0)	&		&	0.50 (D)  \\
	&	C ($Pna2_1$)	&	0.60 (0.57)	&	NT	&	0.38 (D)	& &	C ($Pnma$)	&	0.05 (0.01)	&	90	&	0.74 (D)  \\
	&	E  ($Pna2_1$)	&	0.92 (0.93)	&	NT	&	0.41 (I)	& &	E/C ($MM\_Cmcm$)	&	0.49 (0.47)	&	2050	&	0.32 (D)  \\

\end{tabular}
\end{ruledtabular}}
\end{table*}

\begin{table*}
\caption{\label{rankme2} Ranking of phases by $\Delta$$E$, the total energy per 20-atom cell for the ten $AB$S$_\text{3}$ $B$ = V, Nb, Sc, and Y compounds described in the text. The information below is described in the caption of Table \ref{rankme}.}
\scalebox{0.79}{\begin{ruledtabular}
\begin{tabular}{|c|cccc|ccccc|} 
$A$ & Motif &  $\Delta$$E$ (+ZPE) &  $T_{ trans}$ &$E_g$ & & Motif &  $\Delta$$E$ (+ZPE)  & $T_{ trans}$ & $E_g$  \\
 & (Sp. Grp.) &  (eV/unit cell) &  ( K) &  (eV) & & (Sp. Grp) &  (eV/unit cell)  & ( K) &  (eV)  \\
\hline
& \multicolumn{9}{|c|} {$A$$^\text{$1+$}$$B$$^\text{$5+$}$} \\ 
\hline
 &  \multicolumn{4}{c|} {V} &  & \multicolumn{4}{c|} {Nb} \\	
\hline											
K	&	F ($C222_1$)	&	0 (0)		&		&	0	& &	E ($Sp. Grp. 14$)	&	0 (0)	&		&	0.52 (I)  \\
	&	C ($Pna2_1$)	&	0.23 (0.27)	&	NT	&	0	& &	F ($C222_1$)	&	0.31 (NP)	&	“------”	&	0  \\
	&	E ($Pna2_1$)	&	1.34 (1.31)	&	\textgreater4000	&	0.20 (I)	& &	C ($Pna2_1$)	&	0.47 (0.52)	&	NT	&	0  \\
	&		&		&		&		& &	F/E/C($MM\_Ilmen)$	&	0.54 (0.60)	&	NT	&	1.05 (I)  \\
\hline																		
Rb	&	F ($C222_1$)	&	0 (0)	&		&	0	& &	F ($Sp. Grp. 4$)	&	0 (0)	&		&	0 \\
          &			&			&		&		& &	E ($Pna2_1$)	&	0.26 (0.40)	&	3010	&	0.50 (I)  \\
	&		&		&		&		& &	F/E/C($MM\_Ilmen)$	&	0.60 (0.61)	&	2680	&	0.89 (I)  \\
\hline		
Cs	&	F ($C222_1$)	&	0 (0)	&		&	0	& &	F ($Cmc2_1$)	&	0 (0)	&		&	0.14 (D)  \\
	&	E ($P1$)	&	1.23 (1.26)	&	\textgreater4000	& 0 &	&	E ($Pna2_1$)	&	0.83 (0.80)	&	NT	&	0.61 (I)  \\
	&		&		&		&		& &	F/E/C($MM\_Ilmen)$	&	0.91 (0.92)	&	\textgreater4000	&	0.64 (I)  \\
\hline																		
Tl	&	E ($Pna2_1$)	&	0 (0)	&		&	0	& &	E ($Sp. Grp. 14$)	&	0	(0)	&		&	0.01  \\
	&	F ($C222_1$)	&	0.12 (0.16)	& 1780 	&	0	& &	E/C ($MM\_Cmc2_1$)	&	0.44	(0.44)	&	1960	&	0  \\
	&	E/C ($MM\_Cmc2_1$) & 0.46 (0.46)	& NT	&	0	& &	F ($Cmc2_1$)	&	1.26 (1.28)	&	3475	&	0  \\
	&F/E/C($MM\_Ilmen)$&0.84 (0.82)	& \textgreater4000 	& 0.07 (I) 		& &F/E/C($MM\_Ilmen)$ &1.05 (NP)		&		&	0.36 (I)  \\
\hline																		

 & \multicolumn{9}{|c|} {$A$$^\text{$3+$}$$B$$^\text{$3+$}$} \\ 
\hline
 &  \multicolumn{4}{c|} {Sc} &  & \multicolumn{4}{c|} {Y} \\	
\hline
Bi	&	E ($Pnma$)	&	0 (0)	&		&	1.23 (I)	& &	E ($Pnma$)	&	0	(0)	&		&	1.26 (I)  \\
	&	C ($Pnma$)	&	0.11 (NP)	&	“-----”	&	1.36 (D)	& &	C ($Pna2_1$)	&	0.41 (0.39)	&	\textgreater4000	&	1.72 (I)  \\
	&	E/C ($MM\_Cmc2_1$)	&	0.38 (0.35)	&	1120	&	1.43 (I)	& &	E/C ($MM\_Cmc2_1$)	&	0.42 (0.40)	&	1420	&	1.78 (I)  \\
	&		F/E/C($MM\_Ilmen)$&	0.84 (0.82)	&	2800	&	1.18 (I)	& &	F/E/C($MM\_Ilmen)$	& 0.50 (0.48)		&2275	&	1.49 (I)  \\

\end{tabular}
\end{ruledtabular}}
\end{table*}

Tables \ref{rankme} and \ref{rankme2} rank, for each compound, the lowest energy phases by motif as obtained by DFT where $E$$_\text{$DFT$}$ = $H$($T$ = 0 K).  The phonon assessment at the $\Gamma$-point shows that, except for three compounds, ($A$NbS$_\text{3}$ $A$ = K, Rb, and Tl),  all of the ground state phases are stable with respect to relaxations within the designated  space groups.   For the three cases where stability within the designated space groups was not established, we lifted the space group restriction and  perturbed coordinates to obtain relaxed structures that were evaluated as stable.  These  were slightly lower in energy by at most 0.009 eV/20-atom unit cell as compared to the higher symmetry structure.  These lower symmetry structures maintain the same motif as their higher symmetry parent structures.   Addition of zero-point energies (ZPE) to $E$$_\text{$DFT$}$ does not change the rankings.   The $T$$_\text{trans}$ column indicates the temperature at which the ground state compound and another listed compound have the same free energies as a result of vibrational entropy differences.  For many of these phase transitions, the LDA calculated band gaps for the different phases are significantly different, as can be seen for the compounds with $A$ = Ba and $B$ = Ti and Zr.

\begin{table*}
\caption{\label{batis3} Expanded view of the phases of BaTiS$_\text{3}$.  All energies are with respect to a 20-atom unit cell, which is the number of atoms in the unit cell of the ground state, $C222_1$. NA = phonon frequency/stability not attempted.  All other nomenclature as in Table \ref{rankme}.}
\scalebox{0.75}{\begin{ruledtabular}
\begin{tabular}{|cccc|} 
Motif &  $\Delta$$E$ (+ZPE) & $T_{ trans}$ & $E_g$ \\
 (Sp. Grp.) &  (eV/20-atom unit cell) &  ($^{\circ}$K) &  (eV)  \\
\hline
F ($C222_1$)	&	0.00	(0)	&		&	0	 \\
F ($Cmc2_1$)	&	0.01	(NP)	&	-----	&	0	 \\
F ($P6_3cm$)	&	0.12	(NP)	&	-----	&	0	 \\
F ($P6_3mc$)	&	0.20	(NP)	&	-----	&	0	 \\
F ($P6_3/mmc$)	&	0.22	(NP)	&	-----	&	0	 \\
F ($Cmcm$)	&	0.02	(NP)	&	-----	&	0	 \\
F/C ($2/3$)	&	0.37	(NA)	&	-----	&	0	 \\
F/C ($1/2$)	&	0.37	(NA)	&	-----	&	0	 \\
F/C ($1/3$)	&	0.50	(NA)	&	-----	&	0	 \\
C ($Pna2_1$)	&	0.59	(0.57)	&	NT	&	0.38 (D)	 \\
C ($Pnma$)	&	0.65	(NP)	&	-----	&	0	 \\
C($ R3mR$)	&	0.75	(0.46)	&	340	&	0	 \\
C ($Pm3m$)	&	0.75	(0.47)	&	340	&	0	 \\
C ($P4mm$)	&	0.80	(0.51)	&	375	&	0	 \\
E ($Pna2_1$)	&	0.92	(0.93)	&	NT	&	0.41 (I)	 \\
E ($Pnma$)	&	0.95	(NP)	&	-----	&	0.23 (I)	 \\
																	
\end{tabular}
\end{ruledtabular}}
\end{table*}

As shown in Tables \ref{rankme} and \ref{rankme2}, for those compounds which favor the edge-sharing ground state motif, none preferred the $P1$ phase.  Of particular note, the mixed corner- and edge-sharing phase is often found to have a relative energy between the lowest energy edge-sharing phase and the purely corner-sharing phase, perhaps hinting at a transition path between these two motifs.  For those compounds which prefer the face-sharing ground state motif,   all prefer one of the two orthorhombic phases, $C222_1$ or $Cmc2_1$, over the hexagonal phases, excepting the low symmetry phase for RbNbS$_\text{3}$ which is monoclinic with one unit cell angle equal to 90.67$^\circ$.  As well, of the compounds that prefer the face-sharing motif, only BaTiS$_\text{3}$ has  the mixed motif face- and corner-sharing phases within 1 eV/20-atom unit cell of the ground state phase.  Indeed, BaTiS$_\text{3}$ has the most phases within 1 eV/20-atom unit cell of the ground state.  An expanded view of its phases is shown in Table \ref{batis3}. No other compound has the corner-sharing octahedral $P4mm$, $Pm3m$, and $R3mR$ phases within 1 eV/20-atom unit cell of the ground state.  Finally, absent from Tables \ref{rankme} and \ref{rankme2} is the corner-sharing tetrahedral $Pbcm$ phase.  All compounds evaluated in this phase had energies {\textgreater } 1 eV/20-atom unit cell relative to the ground state.

\begin{table*}
\caption{\label{newt} Ranking of compounds by the standard Goldschmidt factor, $t$, the  ratio of the Pettifor chemical scale values for $A$ and $B$, termed here $Pet$$_\text{$A/B$}$, and a modified $t'$, where $t'$ = $t$$\Delta\chi$(S-$A$)/$\Delta\chi$(O-$A$).  $\chi$ represents the Pauling electronegativity.  $P$$_\text{$A$}$ represents the Pettifor chemical scale value for $A$.}
\scalebox{0.75}{\begin{ruledtabular}
\begin{tabular}{|cccc|ccccc|cccc|} 
$A$ & $B$ & $t$ & Motif &  $A$ &$P$$_\text{$A$}$&$B$ & $Pet$$_\text{$A/B$}$ & Motif  &  $A$ & $B$ & $t'$ & Motif\\ \hline
Bi	&	Y	&	0.831	&	E	&	Bi	&     2.04 &Y	&	0.343	&	E	&	Pb	&	Zr	&	0.207	&	E	\\
Ca	&	Zr	&	0.878	&	EC	&	Bi	&	&Sc 	&	0.363	&	E	&	Pb	&	Ti	&	0.217	&	E	\\
Bi	&	Sc 	&	0.881	&	E	&	Sn	&	1.84 &Zr   &	0.413	&	E	&	Bi	&	Y	&	0.328	&	E	\\
Sn	&	Zr	&	0.884	&	E	&	Pb	&	1.80 &Zr	&	0.422	&	E	&	Bi	&	Sc 	&	0.347	&	E	\\
Sr	&	Zr	&	0.906	&	E	&	Sn	&	1.84 &Ti	&	0.429	&	E	&	Sn	&	Zr	&	0.370	&	E	\\
Ca	&	Ti	&	0.920	&	E	&	Pb	&	1.80 &Ti	&	0.439	&	E	&	Sn	&	Ti	&	0.388	&	E	\\
Pb	&	Zr	&	0.920	&	E	&	Tl	&	1.56 &Nb	&	0.526	&	E	&	Tl	&	Nb	&	0.532	&	E	\\
Sn	&	Ti	&	0.925	&	E	&	Tl	&	&V	&	0.538	&	E	&	Tl	&	V	&	0.555	&	E	\\
Sr	&	Ti	&	0.949	&	E	&	Ca	&	0.60 &Zr	&	1.267	&	EC	&	Ca	&	Zr	&	0.569	&	EC	\\
Ba	&	Zr	&	0.953	&	E	&	Ca	&	&Ti	&	1.317	&	E	&	Sr	&	Zr	&	0.593	&	E	\\
Pb	&	Ti	&	0.963	&	E	&	Sr	&	0.55 &Zr	&	1.382	&	E	&	Ca	&	Ti	&	0.596	&	E	\\
K	&	Nb	&	0.992	&	E	&	Sr	&	&Ti	&	1.436	&	E	&	Sr	&	Ti	&	0.621	&	E	\\
Ba	&	Ti	&	0.998	&	F	&	Ba	&	0.50 &Zr	&	1.520	&	E	&	Ba	&	Zr	&	0.632	&	E	\\
Tl	&	Nb	&	1.009	&	E	&	Ba	&	&Ti	&	1.580	&	F	&	Ba	&	Ti	&	0.661	&	F	\\
Rb	&	Nb	&	1.015	&	F	&	K	&	0.35 &Nb	&	2.343	&	E	&	K	&	Nb	&	0.667	&	E	\\
K	&	V	&	1.034	&	F	&	K	&	&V	&	2.400	&	F	&	Rb	&	Nb	&	0.682	&	F	\\
Tl	&	V	&	1.052	&	E	&	Rb	&	0.30 &Nb	&	2.733	&	F	&	K	&	V	&	0.695	&	F	\\
Rb	&	V	&	1.058	&	F	&	Rb	&	&V	&	2.800	&	F	&	Rb	&	V	&	0.711	&	F	\\
Cs	&	Nb	&	1.061	&	F	&	Cs	&	0.25 &Nb	&	3.280	&	F	&	Cs	&	Nb	&	0.716	&	F	\\
Cs	&	V	&	1.105	&	F	&	Cs	&	&V	&	3.360	&	F	&	Cs	&	V	&	0.747	&	F	\\

\end{tabular}
\end{ruledtabular}}
\end{table*}

In Table \ref{newt}, the standard $t$ for the set of 20 compounds  used in this study is computed, ranked, and compared to the ground state structural motif.  As can be seen from Table \ref{newt}, there are  no overlapping regions of $t$ values for edge- and corner-sharing compounds, simply because there are no compounds which have been calculated to have a corner-sharing ground state.   With the exception of the  Tl-based materials,  the ground state face- and edge- sharing phases can also be predicted using  the standard $t$ factor.  We also define the ``Pettifor factor", $Pet$$_\text{$A/B$}$ = $P$$_\text{$A$}$/$P$$_\text{$B$}$, which we define as the ratio of the Pettifor's chemical scale values for $A$ and $B$ and rank the data accordingly.  In this scenario, the $A$ = Tl compounds are no longer out of line.  Moreover, for the set of compounds chosen for this study, there is a very strong correlation between $P$$_\text{$A$}$ and the ranking of the compounds by $Pet$$_\text{$A/B$}$.  As mentioned in the Introduction, the standard $t$ factor yields overlapping regions of edge- and corner-sharing compounds while the Pettifor chemical scales yields multiple regions of edge- and corner-sharing compounds. The reason ours do not is that existing materials, which have been synthesized at high temperature, are not always created in the ground state, unlike our DFT calculations which determine the ground-state energy at $T$ = 0~K, as will be discussed in the next section.

The predictive ability of the tolerance factor can be further enhanced by taking electronegativity into account in a manner similar to Pearson,\cite{Pearson62p103} and specifying a  new generalized $t$ factor, $t'$ =  $t$$\Delta\chi$($X$-$A$)/$\Delta\chi$(O-$A$), where $\Delta\chi$($X$-$A$) is the electronegativity difference between $X$ = (S, O) and $A$,  and $\Delta\chi$(O-$A$) is the electronegativity difference between O and $A$.   This  results in a ranking of the compounds found in the right side of Table \ref{newt}.  The same ranking would also be found if the denominator of the ratio, $\Delta\chi$(O-$A$), was not included in the formula.  However, by including it, $t'$ remains equivalent to the original $t$ for oxides.  A formulation of $t'$ with a denominator of the ratio set to $\Delta\chi$(F-$A$), would lead to the same ranking again, but now be based on the absolute ranking of electronegativity of the elements in which F has the most negative value. This formulation would be more in the spirit of Pettifor's chemical scale, but it would lose the transferability back to the historic $t$ factor values.  Along these same lines, the $\sqrt{2}$ geometric factor in $t$ is not needed to produce the rankings for either $t$ or $t'$, and it loses its significance in phases that have edge-, face-, and mixed-sharing motifs.  As with all $t$ factors,\cite{Bhalla00p3}  the $t'$ construct is not perfect, as now the ranking of face-sharing BaTiS$_\text{3}$ and edge-sharing KNbS$_\text{3}$ with respect to $t'$ is reversed (but only by 0.006 units).

\section{Discussion}

Several of the 20 compounds considered in our study have been found experimentally to be in a different structural motif phase  than the one we calculated as the ground state phase (BaTiS$_\text{3}$, $A$ZrS$_\text{3}$ with $A$ = Ca, Ba, and the misfit $A$TiS$_\text{3}$ with $A$ = Sn, Pb, and Sr) and others have not yet been synthesized (all $A$$^\text{1+}$$B$$^\text{5+}$S$_\text{3}$ and $A$$^\text{3+}$$B$$^\text{3+}$S$_\text{3}$).  Nevertheless, there is experimental evidence that supports our results, specifically that face-sharing ground state phases have been attained for large $A$ and small $B$ cations, as well as the preponderance of edge-sharing ground state phases for all other $AB$ combinations.

Our calculations show that the ground state phase of  four out of the five $B$ = Zr compounds in this study is the edge-sharing NH$_\text{4}$CdCl$_\text{3}$ $Pnma$ phase.  Only the CaZrS$_\text{3}$ ground state is different, being of mixed edge- and corner-sharing motif and, even in this case, the next higher energy state is predicted to be the edge-sharing NH$_\text{4}$CdCl$_\text{3}$ $Pnma$ phase as well.  
All five of these compounds have been synthesized:  two  in the edge-sharing $Pnma$ phase (PbZrS$_\text{3}$ and SnZrS$_\text{3}$),\cite{Lelieveld78p3348,Yamaoka72p111,Meetsma93p2060} two in the corner-sharing $Pnma$ phase (CaZrS$_\text{3}$ and BaZrS$_\text{3}$),\cite{Lelieveld80p2223, Clearfield63p135} and one in both phases (SrZrS$_\text{3}$).\cite{Lee05p1049}  
Prior theoretical calculations have also shown that the NH$_\text{4}$CdCl$_\text{3}$  $Pnma$ phase is the lowest energy perovskite phase for BaZrS$_\text{3}$.\cite{Bennett09p235115}  Lelieveld {\em et al.}\ and Clearfield have also synthesized SrZrS$_\text{3}$, but only in the edge-sharing phase.\cite{Lelieveld80p2223, Clearfield63p135}   

Our predictions of phase transformations due to small energy differences between the phases provide the insight into the discrepancies between the SrZrS$_\text{3}$ results of Lelieveld {\em et al.}\cite{Lelieveld80p2223} and Clearfield\cite{Clearfield63p135} on the one hand and Lee {\em et al.}\cite{Lee05p1049}  on the other hand.
 In 2005, Lee {\em et al.}\ synthesized  edge-sharing SrZrS$_\text{3}$ by mixing the constituent elements together in stoichiometric proportions and then heating at 1120 K.  Performing the same procedures at 1220 K led to the creation of a two-phase material with a major corner-sharing phase and a minor edge-sharing phase.\cite{Lee05p1049}  In 1980, Lelieveld {\em et al.} flowed H$_\text{2}$S gas over mixtures of binary oxides at 1370 K to create a solely corner-sharing perovskite.\cite{Lelieveld80p2223}    Thus, it is probably the differences in processing temperatures and starting materials that led to the different results between these two experiments. Interestingly, in 1963, Clearfield explored the effect of temperature on the synthesis of BaZrS$_\text{3}$, SrZrS$_\text{3}$, and CaZrS$_\text{3}$  in a manner similar to that of Lelieveld {\em et al.}  In Clearfield's method, he first combined binary oxides to form $A$ZrO$_\text{3}$, then used CS$_\text{2}$ gas to replace O with S.  He discovered that for synthesis temperatures between 1020 - 1270 K, an unknown phase of BaZrS$_\text{3}$ was present in sizable amounts (10-15\% composition of the product); and, for all three, at temperatures below 1270 K, found it impossible to state the space group with certainty.\cite{Clearfield63p135}  Based on the work of  Lee {\em et al.}, we suggest that it is possible that Clearfield obtained both the corner- and edge-sharing structures within each composition. 

In our theoretical study, we show that for SrZrS$_\text{3}$ the corner-sharing phase is preferred at temperatures above 2050~K.  With respect to Lee's results, our transition temperature at which the corner-sharing phase is preferred over the edge-sharing phase is approximately 900~K too high.   Though we do calculate a phase change near 1200~K, it is for a change to a mixed motif corner- and edge-sharing one, and not a completely corner-sharing phase.  Thus, as our calculation method involves only harmonic $\Gamma$-point phonon contributions to energy, our errors can be attributed to not including full Brillouin zone averaging and anharmonic energy contributions.  For PbZrS$_\text{3}$ and SnZrS$_\text{3}$, the phase changes from edge-sharing to corner-sharing have similar crossover temperatures to SrZrS$_\text{3}$.  As they have been synthesized as edge-sharing phases at 1070 K,\cite{Lelieveld78p3348,Yamaoka72p111,Meetsma93p2060}  our study suggests that they can also be made as corner-sharing phases by synthesizing at higher temperatures.

For BaZrS$_\text{3}$, we calculate that the edge-sharing phase is energetically preferred below 90 K.  As temperatures in the vicinity of 90 K are too low for synthesis, it would seem that, by itself, a change in synthesis temperature will not lead to the formation of the edge-sharing phase.   While CaZrS$_\text{3}$ has only been made in the corner-sharing phase, we have shown that this phase is not energetically preferred over the edge-sharing phase or the mixed motif phase at any temperature.   Therefore, it should be possible to achieve these other phases of CaZrS$_\text{3}$ through either lower synthesis temperatures alone or in combination with other changes in synthesis procedures such as increased pressure.  Supporting this idea is the existence of  another $ABX_3$ compound with $A$ = Ca, CaIrO$_\text{3}$, which is created in the mixed  edge- and corner-sharing phase through the use of elevated pressures.\cite{Martin07p1912}

Next, we compare our theoretical space group and structure predictions of stoichiometric ternary sulfides  in which the $A$ cations have a lone pair electron configuration, (and with $B$ not equal to Zr), with experimental literature for those systems where non-stoichiometric phases are reported.  For $AB$S$_\text{3}$ with $A$ cations that possess a lone pair, (Pb, Sn, and Bi), the nonstoichiometric phases are chiefly composed of single sheets of edge-sharing $B$S$_\text{6}$ octahedra with a chemical formula of $B$S$_\text{2}$ separated by single or multiple planes of distorted rock salt $A$S.\cite{Wiegers92p351} An important point of agreement is that our calculations also predict an edge-sharing structure for each of these. However, the stoichiometric phases prefer pairs of columns of edge-sharing octahedra, rather than the sheets seen in the misfit compounds.  Despite this difference, there is  experimental evidence that these two findings are compatible.
Wiegers and Meerschaut synthesized  (LaS)$_\text{1+$x$}$$B$S$_\text{2}$  ($B$ = Ti, V, and Cr)  misfits under atmospheric pressure conditions.\cite{Wiegers92p351}  Kikkawa {\em et al.} formed stoichiometric La$B$S$_\text{3}$ in the  edge-sharing NH$_\text{4}$CdCl$_\text{3}$ phase for the same $B$ species by applying high pressure to the mixture of reactants.\cite{Kikkawa98p233} 
Thus, the elevated pressure synthesis method of Kikkawa {\em et al.}\ is probably necessary for the stoichiometric $AB$S$_\text{3}$ formation of the systems containing $A$ = Pb, Sn, and $B$ = Ti.

The compounds in our study in which the $A$ cations have a lone pair configuration, (and with $B$ not equal to Zr), also have another common feature.  Our calculations show that several of them transition from the edge-sharing phase to the mixed motif edge- and corner-sharing phases at similar temperatures: 1150 K, 1120 K, and 1420 K for PbTiS$_\text{3}$, BiScS$_\text{3}$, and BiYS$_\text{3}$ respectively.  TlNbS$_\text{3}$ has the same phase transition at a higher temperature, 1960 K.  As well, CaTiS$_\text{3}$, which does not have a lone pair for $A$ = Ca, also exhibits the potential for this transformation, at 305 K. 

In order to evaluate our results for the six compounds which are found to have the face-sharing motif as the lowest energy phase, we separate them into two groups:  five with Group I $A$ = (Cs, Rb, and K), which have not been made experimentally in any phase, and BaTiS$_\text{3}$, which has been synthesized by multiple research groups.  The face-sharing ground state of these Group I compounds is similar to the related face-sharing structures of $AB$Cl$_\text{3}$, with different $B$.  They do not have  the $Pbcm$ structural motif of their $AB$O$_\text{3}$ analogs, (single columns of corner-sharing tetrahedra for $B$ = V), nor the double columns of edge-sharing octahedra for $B$ = Nb.  This indicates that the face-sharing motif is not only a function of  size of  Group I $A$, but a function of the $X$ size as well:  both Cl$^\text{-}$ and S$^\text{2-}$ are very large and quite close in size  in different environments: their ionic radii are 1.81 and 1.84 \AA\ respectively, when adopting a coordination of 6;\cite{Shannon76p751,Seshadri11p1} and their covalent radii are 0.99 and 1.02 \AA\ respectively.\cite{Sanderson62p1}   A notable difference between the sulfides and their chloride analogs is that the latter form mostly in a hexagonal  lattice, rather than an orthorhombic one.  An interesting exception is RbCrCl$_\text{3}$, which is nearly orthorhombic at room temperature with a monoclinic classification and an angle deviation from 90$^{\circ}$ of $\approx$3$^{\circ}$.\cite{Zandbergen81p199}  In conjunction with this observation, we have found that the lowest energy phase for RbNbS$_\text{3}$ is also monoclinic albeit with a smaller angle deviation from 90$^{\circ}$, 90.67$^{\circ}$.

For the lone Group II $A$ that assumes a face-sharing ground state, BaTiS$_\text{3}$, we find that it prefers the orthorhombic C222$_\text{1}$ space group.  In the experimental literature, on the other hand, it is listed in one of  two  hexagonal space groups,  $P6$$_\text{3}$$/mmc$ or $P6_3mc$.\cite{Aslanov64p1317,Huster80p775,Clearfield63p135}  However, Clearfield has noted  that at lower temperatures of synthesis ($\approx$970 K), the compound could be characterized with either orthorhombic or hexagonal indexing.\cite{Clearfield63p135} As the synthesis temperature was increased to 1370 K, only hexagonal characterization was plausible.  Thus, similarly to  SrZrS$_\text{3}$, the structure of BaTiS$_3$ is sensitive to changes in the synthesis temperature.  Further, both Fagot {\em et al.} and Ghedira {\em et al.} have shown experimentally that an analog of BaTiS$_\text{3}$, BaVS$_\text{3}$, undergoes a phase change from hexagonal  $P6$$_\text{3}$$/mmc$ phase to an orthorhombic phase (either $Cmc2_1$ or $C222_1$) when temperature is lowered below $\approx$250 K.\cite{Fagot05p718,Ghedira81p1491}  Since our DFT calculations are performed at 0~K, our BaTiS$_\text{3}$  results are consistent with their findings and also explain the calculated preferred orthorhombic phases as opposed to  hexagonal phases for the Group I face-sharing compounds.  Fagot {\em et al.} proposed that BaVS$_\text{3}$ changes to a $C1m1$ phase as the temperature is lowered below 70~K.\cite{Fagot05p718}  To test whether this phase was possible for BaTiS$_3$, we performed a relaxation of BaTiS$_\text{3}$ assuming the $C1m1$ phase and found that it was slightly lower in energy ($\Delta$E {\textless } 0.002 eV/20-atom unit cell) than the previously calculated  $C222_1$ ground state; however,  the $\Gamma$-point phonon calculation showed that this phase was not stable at 0~K.  
These analyses also demonstrate one limitation of our work:  when many phases are similar in energy, our free energy approximation can reorder the phases.

As is shown in Table \ref{batis3}, the mixed motif face- and corner-sharing phases for BaTiS$_\text{3}$   energy levels fall between the  wholly face-sharing phases and the wholly corner-sharing phases.  The compound's oxide analog, BaTiO$_\text{3}$, is most often cited to be a corner-sharing phase compound.   It is worthy of note that BaTiO$_\text{3}$ has also been processed in the $1/3$ phase, which is more formally known as the BaFeO$_\text{2+x}$ phase.\cite{Akimoto94p160, Burbank48p330, Yashima05p014313}  Thus, our results are consistent with the literature analogs.   More importantly though, our calculations indicate that phase changes are possible from face- to corner-sharing motifs, as we calculate transition temperatures from the $C222_1$ phase to the $R3mR$, $Pm3m$, and $P4mm$ phases in the 340-375 K range.   Based on the DFT calculations, we propose that BaTiS$_\text{3}$ will  be found to be a highly structurally flexible material when synthesized by different experimental methods.

\section{Conclusions}

From a set of 22 phases known for $ABX$$_\text{3}$ compounds, we found that, for  $AB$S$_\text{3}$ compounds in which the $B$ element has a $d$$^\text{$0$}$ electronic configuration, the preferred phase for all but the largest $A$ cations and smallest $B$ cations are the edge-sharing $Pna2_1$ and NH$_\text{4}$CdCl$_\text{3}$ $Pnma$ phases.  These sulfides differ from their oxide counterparts, which favor corner-sharing phases.  To predict the preferred structural motifs, we developed a modified Goldschmidt tolerance factor $t'$.  This  incorporates the electronegativity difference between the $A$ cation and S, but retains the original $t$ for oxides, by normalizing the difference in electronegativity  between the $A$ cation and O.  This formulation leads to a neat demarcation between the compounds that prefer a face-sharing ground state and those that prefer an edge-sharing one.  

Several of the $AB$S$_\text{3}$ combinations have phases with different motifs that are within 1 eV/20-~atom unit cell of the energy of the ground state phase.  Vibrational entropy calculations show that  these phases might be achievable under different synthesis conditions than the ones already present in the literature.  For the smaller $A$ and $B$ cation combinations in $AB$S$_\text{3}$, high synthesis temperatures under ambient pressure conditions, often with oxide intermediates or binary oxide starting materials, have led to products with the corner-sharing motif forming or to incommensurate phases.  Experimental evidence shows that combinations of high pressure, lower processing temperatures, non-oxide starting materials, and long processing times tend to favor the synthesis of the commensurate edge-sharing motif.  
We  suggest that two of the sulfides that are evaluated in this paper, (BaZrS$_\text{3}$ and CaZrS$_\text{3}$),  are candidate compounds that may be produced as edge-sharing phases in this manner.  Conversely, though PbZrS$_\text{3}$  and SnZrS$_\text{3}$ have been synthesized as edge-sharing compounds, higher synthesis temperatures could produce corner- and mixed motif corner- and edge-sharing phases.  Lastly, BaTiS$_\text{3}$  might achieve both hexagonal and orthorhombic face-sharing motifs and corner-sharing motifs when subjected to different synthesis temperatures.  Thus, not only does the family of $AB$S$_\text{3}$ compounds show structural diversity, but even the individual $AB$S$_\text{3}$ compounds themselves exhibit structural diversity with multiple stable phases.

\section{Acknowledgments}
JAB was supported by the Office of Naval Research, under grant N00014-12-1-1033.  JWB and MRS were supported by the AFOSR, under grant FA9550-10-1-0248.  IG was supported by the National Science Foundation, under grant DMR11-24696.  AMR was supported by the Department of Energy Office of Basic Energy Sciences, under grant number DE-FG02-07ER46431.  Computational support was provided by the HPCMO of the U.S. DoD and the NERSC of the DoE.


\begin{thebibliography}{10}

\bibitem{Muller74p1}
O. ~Muller and R. Roy,
\newblock $The Major Ternary Structural Families$, Springer-Verlag, New York, NY, USA (1974).


\bibitem{Clearfield63p135} 
A. ~Clearfield,
\newblock Acta Cryst. {\bf 16} 135 (1963).

\bibitem{Belsky02p364}
A. ~Belsky, M. ~Hellenbrandt, V. ~L.  ~Karen, and P. ~Luksch,
\newblock Acta Cryst. B {\bf 58} 364 (2002).

\bibitem{FIZ13p1}
FIZ Karlsruhe ICSD Database Data Release 2013.2.

\bibitem{Goldschmidt26p477}
V. ~M. Goldschmidt,
\newblock Die Naturwissenschaften {\bf 14} 477 (1926).

\bibitem{Woodward97p44}
P. ~M. ~Woodward, 
\newblock Acta Cryst. {\bf B53} 44 (1997).


\bibitem{Goodenough04p1915}
J. ~B. ~Goodenough,
\newblock Rep. Prog. Phys. {\bf 67} 1915 (2004).

\bibitem{Pettifor84p31}
D. ~G. ~Pettifor,
\newblock Sol. St. Comm. {\bf 51} 31 (1984).

\bibitem{Ferro96p205}
R. ~Ferro and A. ~Saccone,
\newblock $Physical Metallurgy$, 4th Ed., edited by  R. ~W. ~Cahn and P. ~Haasen, (Elsevier Science, Amsterdam, The Netherlands, 1996), Vol. I, Chap. 4.

\bibitem{Pettifor88p675}
D. ~G. ~Pettifor,
\newblock Mat. Sci. and Tech.  {\bf 4} 675 (1988).

\bibitem{Lee05p1049}
C.-S. ~Lee, K. ~M. ~Kleinke, H. ~Kleinke,
\newblock Sol. St. Sci. {\bf 7} 1049 (2005).


\bibitem{Gonze02p478}
X. ~Gonze, J.-M. ~Beuken, R. ~Caracas, F. ~Detraux, M. ~Fuchs,
G.-M. ~Rignanese, L. ~Sindic, M. ~Verstraete, G. ~Zerah, F. ~Jollet,
M. ~Torrent, A. ~Roy, M. ~Mikami, Ph. ~Ghosez, J.-Y. ~Raty, D. ~C. ~Allan,
\newblock Comput. Mater. Sci. {\bf 25} 478 (2002).



\bibitem{Glazer72p3384}
A. ~M. ~Glazer,
\newblock Acta Cryst. B {\bf 28} 3384 (1972).



\bibitem{Fagot05p718}
S. ~Fagot, P. ~Foury-Leylekian, S. ~Ravya, J.~P. ~Pouget, M. ~Annec, G. ~Popov, M. ~V. ~Lobanov, M. ~Greenblatt,
\newblock Sol. St. Sci. {\bf 7} 718 (2005).

\bibitem{Ghedira81p1491}
M. ~Ghedira, J. ~Chenavas, F. ~Sayetat, M. ~Marezio, O. ~Massenet, and J. ~Mercier,
\newblock Acta Cryst. {\bf B37} 1491 (1981).


\bibitem{Murakami04p855}
M. ~Murakami, K. ~Hirose, K. Kawamura, N. ~Sata, and Y. ~Ohishi,
\newblock Science {\bf  304} 855 (2004).

\bibitem{Martin07p1912}
C. ~D. ~Martin, R. ~I. Smith, W. ~G. ~Marshall, and J. ~B. ~Parise,
\newblock Amer. Mineralogist {\bf 92} 1912 (2007).


\bibitem{Ramer99p12471}
N. ~J. ~Ramer and A. ~M. ~Rappe, 
\newblock Phys. Rev. B {\bf 59} 12471 (1999).

\bibitem{Rappe90p1227}
A. ~M. ~Rappe, K. ~M. ~Rabe, E. ~Kaxiras, J. ~D. ~Joannopoulos,
\newblock Phys. Rev. B {\bf 41} 1227 (1990).

\bibitem{OpiumSourceforgenet11p1} 
http://opium.sourceforge.net.



\bibitem{Monkhorst76p5188}
H. ~J. ~Monkhorst and J. ~D. ~Pack,
\newblock Phys. Rev. B {\bf 13} 5188 (1976).







\bibitem{Seshadri11p1}
R. ~Seshadri,
\newblock http://www.mrl.ucsb.edu/~seshadri/Periodic/index.html.


\bibitem{Shannon76p751}
R. ~D. ~Shannon,
\newblock Acta Cryst. {\bf A32} 751 (1976).

\bibitem{Bennett11p144112}
J. ~W. ~Bennett, I. ~Grinberg, P. ~K. ~Davies, and A. ~M. ~Rappe,
\newblock Phys. Rev. B {\bf 83} 144112 (2011).



\bibitem{Pearson62p103}
W. ~B. ~Pearson,
\newblock J. Phys. Chem. Solids {\bf 23} 103 (1962).

\bibitem{Bhalla00p3} 
A. ~S. ~Bhalla, R. ~Guo, and R. ~Roy,
\newblock Mat. Res. Innovat. {\bf 4} 3 (2000).



\bibitem{Lelieveld78p3348}
R. ~L. Lelieveld and D. ~J. ~W. ~Ijdo,
\newblock Acta Cryst. {\bf B34}  3348 (1978).

\bibitem{Yamaoka72p111}
S. ~Yamaoka,
\newblock J. Of Amer. Cer. Soc. {\bf 55} 111 (1972).

\bibitem{Meetsma93p2060}
A. ~Meetsma, A. ~Wiegers, and J. ~L. ~DeBoer,
\newblock Acta Cryst. {\bf C49} 2060 (1993).


\bibitem{Lelieveld80p2223}
R. ~Lelieveld and D. ~J. ~W. ~Ijdo,
\newblock Acta Cryst. {\bf B36} 2223 (1980).



\bibitem{Bennett09p235115}
J. ~W. ~Bennett, I. ~Grinberg, A. ~M. ~Rappe,
\newblock Phys. Rev. B {\bf 79} 235115 (2009).

\bibitem{Wiegers92p351}
G. ~A. Wiegers and A. ~Meerschaut,
\newblock J. of Alloys and Comp. {\bf 178} 351 (1992).


\bibitem{Kikkawa98p233}
S. ~Kikkawa, Y. ~Fujii, Y. ~Miyamoto, F. ~Kanamaru, A. ~Meerschaut, A. ~Lafond, and J. ~Rouxel,
\newblock J. of Sol. St. Chem. {\bf 139} 233 (1998).


\bibitem{Sanderson62p1}
R. ~T. Sanderson,
\newblock $Chemical Periodicity$, Reinhold Publishing Corp., New York, NY, USA (1962).


\bibitem{Zandbergen81p199}
H. ~W. ~Zandbergen and D. ~J. ~W. ~Ijdo,
\newblock J. of Sol. St. Chem. {\bf 38} 199 (1981)

\bibitem{Aslanov64p1317}
L. ~A. ~Aslanov and L. ~M. ~Korba,
\newblock Russ. J. Inorg. Chem. {\bf 9}, 1317 (1964).

\bibitem{Huster80p775} 
J. ~Huster
\newblock Zeitshrift for Naturforschung {\bf 35}, 775 (1980).

\bibitem{Akimoto94p160}
J. ~Akimoto, Y. ~Gotoh, and Y. ~Oosawa,
\newblock Inorganic Compounds C {\bf50}, 160, (1994).

\bibitem{Burbank48p330}
R. ~D. Burbank and H. ~T. Evans, Jr.,
\newblock Acta. Cryst. {\bf 1}, 330 (1948)

\bibitem{Yashima05p014313}
M. ~Yashima, T. ~Hoshina, D. ~Ishimura, S. ~Kobayashi, W. ~Nakamura, T. ~Tsurumi, and S. ~Wada,
\newblock J. of App. Phys. {\bf 98} 014313 (2005).


\end{thebibliography}
\end{document}